\documentclass[manuscript]{aastex}

\slugcomment{Not to appear in Nonlearned J., 45.}

\shorttitle{Ages and Metallicities in A779 with narrowband continuum colors}
\shortauthors{Sreedhar et al.}

\begin{document}


\title{Ages and Metallicities of Cluster Galaxies in A779 using Modified Str\"{o}mgren Photometry}


\author{Yuvraj Harsha Sreedhar\altaffilmark{}, Andrew P. Odell\altaffilmark{2}, Karl D. Rakos\altaffilmark{}, Gerhard Hensler\altaffilmark{}, Werner W. Zeilinger\altaffilmark{}}
\affil{University of Vienna, Institute of Astronomy, T\"{u}rkenschanzstra\ss e 17, A-1180 Vienna, Austria}
\affil{Department of Physics and Astronomy, Northern Arizona University, Flagstaff, Arizona, USA}

%\author{C. D. Biemesderfer\altaffilmark{4,5}}
%\affil{National Optical Astronomy Observatories, Tucson, AZ 85719}
%\email{aastex-help@aas.org}

%\and

%\author{R. J. Hanisch\altaffilmark{5}}
%\affil{Space Telescope Science Institute, Baltimore, MD 21218}


%\altaffiltext{1}{Visiting Astronomer, Cerro Tololo Inter-American Observatory.
%CTIO is operated by AURA, Inc.\ under contract to the National Science
%\altaffiltext{}{}
%\altaffiltext{3}{present address: Center for Astrophysics,
%   60 Garden Street, Cambridge, MA 02138}
%\altaffiltext{4}{Visiting Programmer, Space Telescope Science Institute}
%\altaffiltext{5}{Patron, Alonso's Bar and Grill}


\begin{abstract}
In the quest for the formation and evolution of galaxy clusters, Rakos and co-workers introduced a spectrophotometric method using the modified Str\"omgren photometry. But with the considerable debate toward the project's abilities, we re-introduce the system by testing for the repeatability of the modified Str\"{o}mgren colors and compare them with the Str\"omgren colors, and check for the reproducibility of the ages and metallicities (using the Principle Component Analysis (PCA) technique and the GALEV models) for the six common galaxies in all the three A779 data sets. As a result, a fair agreement between two filter systems was found to produce similar colors (with the precision of 0.09 mag in \textit{(uz-vz)}, 0.02 mag in \textit{(bz-yz)}, and 0.03 mag in \textit{(vz-vz)}) and the generated ages and metallicities are also similar (with the uncertainty of 0.36 Gyr and 0.04 dex from the PCA and 0.44 Gyr and 0.2 dex using the GALEV models). We infer that the technique is able to relieve the age-metallicity degeneracy by separating the age effects from the metallicity effects, but still unable to completely break. We further extend this paper to re-study the evolution of galaxies in the low mass, dynamically poor A779 cluster, (as it was not elaborately analyzed by Rakos and co-workers in their previous work) by correlating the luminosity (mass), density, radial distance with the estimated age, metallicity, and the star formation history. Our results distinctly show the bimodality of the young, low-mass, metal-poor population with the mean age of 6.7 Gyr ($\pm$ 0.5 Gyr) and the old, high-mass, metal-rich galaxies with the mean age of 9 Gyr ($\pm$ 0.5 Gyr). The method also observes the color evolution of the blue cluster galaxies to red (Butcher-Oemler phenomenon), and the downsizing phenomenon. Our analysis shows that the modified Str\"omgren photometry is very well suited for studying low- and intermediate-z clusters, as it is capable of observing deeper with better spatial resolution at spectroscopic redshift limits, and the narrowband filters estimate the age and metallicity with lesser uncertainties compared to other methods that study stellar population scenarios.
\end{abstract}





\section{Introduction}

The initial use of the Str\"omgren filter system was to study stars by Str\"omgren (1966), where one could exploit the filter's sensitive nature to detect individual stellar parameters. However, Rakos et al. (1988) used these filters, with slight modifications, to analyze galaxy evolution in clusters at their rest-frame. Using this unique feature of rest frame Str\"omgren photometry, we compare and analyze the results of A779 cluster galaxies. The cluster A779 has been studied by many authors (Kellogg  et al. 1973; Huchra et al. 1992; White et al. 1997; Jones \& Forman 1984; David et al. 1993; Zabludoff et al. 1993; Hwang \& Lee  2008) in different wavelengths and using different techniques which presented intriguing results on the cluster's formation and evolution. Previously, Rakos et al. (2008) had cumulatively studied eight clusters, including A779. Since its evolutionary properties were significantly overshadowed by the other rich clusters in their study (\objectname{Fornax}, \objectname{Coma}, \objectname{A1185}, \objectname{A2218}, the X-ray bright clusters like \objectname{A539}, {A119} and the X-ray, radio bright cluster \objectname{A400}), we investigate the formation and evolution of galaxies in \objectname{A779}, uniquely.

Also, since the rest-frame narrowband galaxy analysis technique along with its abilities are often debated in the scientific community, therefore a thorough testing of this method becomes mandatory. In order to do so and to show its robustness, we have carried out a repeatability test of the observed \objectname{A779} galaxy colors obtained from two different observations and compared it with the previously published results of Rakos et al. (2008). The three observations compared here are conducted using two filter systems, Str\"omgren and modified Str\"omgren photometry (defined ahead). 

Primary parameters, like the ages and metallicities of galaxies reflect the underlying star formation history and the chemical evolution of galaxies, respectively. But their measurements and reliability are often model dependent and inaccurate. In addition, the age-metallicity degeneracy (Worthey 1994, 1999) further affects their measurements, caused by the contribution of the main-sequence turnoff stars, that are sensitive to the age, and the red giant branch (RGB) stars, which are sensitive to the metallicity. In order to relieve this degeneracy, Rakos et al. (1996) took advantage of the narrowband filters that focus only on certain regions of the galaxy spectrum which are either sensitive to age or metallicity. To further isolate these two parameters, Rakos \& Schombert (2005a, henceforth R05a) implemented the Principle Component Analysis (PCA; Steindling et al. 2001) technique on the Schulz et al.'s (2002, henceforth S02) evolutionary models. As an evaluation, we compare the ages and metallicities (obtained from the observed \objectname{A779} galaxy colors) determined from the PCA method with the advanced GALEV (Kotulla et al. 2009) evolutionary models. 

Using our latest observations of 2009 December, we show the ability of our modified Str\"omgren photometry by observing a poor (Bautz-Morgan, BM class II) and faint (with galaxy members of $M_{5500}$ $\simeq$ $-$14.73) \objectname{A779} galaxy cluster and explore its evolutionary scenarios. It is interesting to study these small, poor clusters, since they resemble the early clusters at high-$z$ that display the accumulation of galaxies to form large clusters. By studying these low mass, poor clusters using modified Str\"omgren photometry, we hope to discover and study some phenomena that may have been overlooked by other disparate studies. Therefore, we consider studies from other wavelengths to understand the cluster's formation and evolution predominantly.

Observations in the X-rays complement the cluster studies greatly by showing the underlying intracluster medium and hot gas, which forms the second massive cluster component, after the dark matter ($\sim$ 80\% cluster mass). Galaxies and the intergalactic medium (IGM) are known to be well connected by the accretion (on to galaxies) and flows of matter and energy by the process of "Cosmic Feedback". Starburst galaxies expel matter by galactic-scale outflows or "superwinds" (originating from hot massive stars and core-collapse supernovae) to the IGM, polluting them with the newly synthesized metals from stars. A part of the lost material is recycled directly within the intergalactic environment, forming the tidal dwarf galaxies (TDGs) that are made out of tidally pulled out colliding galaxies. These TDGs that are commonly found near the interacting systems (Weilbacher et al. 2000), like, in groups and clusters, are formed by the unstable tidal HI content that collapses, transforming into H$_2$  gas and igniting the star formation (Duc et al. 2002). On the other hand, galaxies rotating through the hot IGM, depending upon their mass and morphology, experience the ram-pressure stripping (Gunn \& Gott 1972) that removes their ISM
resulting in either low or no star formation.

\objectname{A779} ($z$=0.023) has been noted for its small size (75\arcmin\ in radius, resembles almost like a galaxy group) with the BM class II and the Abell richness class of 0 has a faint X-ray luminosity showing the presence of a very small amount of the X-ray emitting gas. Table 1 presents \objectname{A779}'s X-ray luminosity measurements by different authors from the literature till today. All these measurements present a low X-ray emission indicating smaller gas content with the gas temperature of 1.5 keV (White et al. 1997). Hwang \& Lee's (2008) \objectname{A779} study using Sloan Digital Sky Survey (SDSS) and 2dFGRS also detect very faint X-ray luminosity along with the mean galaxy radial velocity dispersion of $491^{+40}_{-36}$ km s$^{-1}$. Consistent with the studies of Oegerle \& Hill (2001), Hwang \& Lee (2008) also observe a cD galaxy, \objectname{NGC 2832}, at the center of A779 which is at rest in the cluster potential and the presence of a large number of late-type galaxies compared to the early-types.

On the other hand, radio observations from the Westerbrock survey of cluster galaxies by Wilson \& Vall\'ee (1982) measure the peak map flux density for \objectname{A779} to be 9 mJy. The two radio sources found at the cluster center, \objectname{NGC 2832} and \objectname{NGC 2831} (extended towards the NW-SE), form a double system in a halo. The direction of the radio source elongation seems similar to the optical halo. The core is dominated by the S0s and compact ellipticals, but no X-ray emission was detected by HEAO-2 and no radio emission at the flux level of $\sim$ 5 mJy (Fanti et al. 1984). A more recent 21 cm study by Solanes et al. (2001) infers that the central ($r \leq$ 1 Abell radius) and outer ($r \geq$ 1 Abell radius) spiral population of \objectname{A779} have statistically significant reduced HI content. The peculiarity of this cluster comes from the HI deficient components that show no statistically significant trend with cluster properties, such as the X-ray luminosity, temperature, velocity dispersion and richness of the spiral fraction, which could be due to the higher rate of transformation of swept spirals into lenticulars in the richest X-ray clusters. The amount of gas depletion appears to be related to the morphology of the disks, but it is hardly a function of their optical size (Solanes et al. 2001). 

With the view to test the method and to investigate the formation and evolution of galaxies in A779, the paper is arranged in the following sections. (1) The observation section describes the definition and the continuum colors of the modified Str\"omgren photometry. (2) The results elaborate on the method for estimating galaxy ages and metallicities, and present the comparison of colors, model ages and metallicities. (3) The discussion section reports the evolution of A779 member galaxies using correlations of different galaxy parameters. (4) Conclusion. 

\section{Observation}

\subsection{Modified Str\"omgren photometry: The Definition}

In the literature there are several definitions of the modified Str\"omgren filter system, most of which are not accurately defined. In principle, the modified Str\"omgren \textit{(uz,vz,bz,yz)} filters are not much different from the normal Str\"omgren \textit{(u,v,b,y)} filters, as the effective wavelength of the two filter systems is almost same. However, the modifications in the modified Str\"{o}mgren filters come from the following aspects. (1) The \textit{uz} filter is slightly (30\AA) shifted to the red to focus more on the red galaxies. (2) The calibration of these filters is performed using spectrophotometric standard stars that are shifted to the redshift of the target cluster (Rakos et al. 1988). This makes the zero point of the magnitudes to differ along with the cluster's redshift. (3) The flux measurement for the galaxies and the standard stars is in the wavelength units (per \AA), as usually done for most galaxy studies instead of the frequency (per Hz) units, as in the stellar studies. 

\subsection{\textit{Narrowband Continuum Colors}}

The first set (2004 December) of A779 observation was obtained at 2.3m Bok telescope of the Steward Observatory which offers a $90\arcsec$ prime image device with a prime-focus wide field imager and four 4k $\times$ 4k mosaic CCDs that observed a field of view of 1$^{\circ}$.16 $\times$ 1$^{\circ}$.16 and a plate scale of $0\arcsec.45 pixel^{-1}$ (Rakos et al. 2008). The observational attempts made in 2009 were to repeat just like the 2004 observations, but instead of four 4k $\times$ 4k mosaic CCDs only single 4k x 4k CCD was used with field of view of $30\arcmin$ $\times$ $30\arcmin$ and a plate scale of $0\arcsec.45 pixel^{-1}$. The observation at La Palma was performed using 2.54 m INT prime-focus imager that has a field of view of $20\arcmin$ $\times$ $40\arcmin$ and the wide field camera consists of 2k $\times$ 4k mosaic CCDs and plate scale of $0\arcsec .33pixel^{-1}$.

The filter system used for the 2004 and 2009 observations at Steward Observatory is the same modified Str\"omgren system. Three exposures of 600 s in the \textit{uz} and \textit{vz} and four exposures of 300 s were made in \textit{bz}, and \textit{yz} for the 2004 observations. However, for the latest 2009 observations, three images of 600 s in \textit{uz} and \textit{vz}, three of 300 s in \textit{bz} and four of 300 s in \textit{yz} were obtained. The calibration was obtained through a number of spectrophotometric standards measured on each night for the 2004 set of observations, but for 2009 observations the spectrophotometric standard measurements of four nights were averaged. The modified Str\"omgren filter system covers three regions in the near-UV and blue portion of the electromagnetic spectrum. The \textit{bz} \textit{($\lambda_{eff}$=}$4675$\AA) and \textit{yz} \textit{($\lambda_{eff}$=}$5500$\AA) focus on the continuum part of the spectrum, both of which combine to form the temperature-color index \textit{(bz-yz)}. The \textit{vz} \textit{($\lambda_{eff}$=}$4100$\AA) filter is strongly influenced by the metal absorption lines (i.e., Fe, CN) from the old stellar populations, whereas \textit{uz} \textit{($\lambda_{eff}$=} $3500$\AA) is shortward of the Balmer jump. All the filters are sufficiently narrow (FWHM $\sim$ $200$\AA) to sample the regions of spectrum that are unique to the determination of various physical parameters, like, star formation rate, age, and metallicity (Rakos et al. 2001). For the 2005 INT observation, the standard \textit{(uvby)} filters were used and retained for analysis (by applying proper $k$-corrections to compare with the modified Str\"omgren filter system), wherein five exposures of 600 s in each filter were made. Calibration was performed using the average of five nights of spectrophotometric standards. 

The reduction procedure has been published in Rakos et al. (1996). The photometric system is based on the theoretical transmission curves of filters (which can be obtained upon request from the authors) and the spectra of the spectrophotometric standards are published by Massey \& Gronwall (1990) and Hamuy et al. (1994). The convolution of the transmission curves and the spectra of the standard stars produces theoretical flux values for color indices of the standard stars corrected for all light losses in the equipment and the specific sensitivity of the CCD camera. Magnitudes are measured on the resultant co-added images using standard IRAF procedures and are based for brighter objects on metric apertures set at 32 kpc for cosmological parameters of H$_o$ = 75 km s$^{-1}$ Mpc$^{-1}$ and the benchmark cosmology ($\Omega_m = 0.3$, $\Omega_\Lambda = 0.7$). For the fainter objects, the apertures are adapted to deliver the best possible signal to noise ratio, but always using the same aperture for all four filters.

For the calculated colors, the galactic extinction toward \objectname{A779} is found to be small of \textit{E(B-V)} = 0.017 mag, that corresponds to \textit{E(uz-vz)} = 0.008, \textit{E(bz-yz)} = 0.012 and \textit{E(vz-yz)} = 0.019. The effect of these corrections is negligible on the measured ages and metallicities (shown in Table 2). The limiting magnitude for all the three observations is $13.05 \leq m_{5500} \leq 20.35$. The typical errors in colors for the three observations are in the range: 0.03 $\leq$ $(uz-vz)$ $\leq$ 0.09, 0.02 $\leq$ $(bz-yz)$ $\leq$ 0.05, 0.03 $\leq$ $(vz-yz)$ $\leq$ 0.07. Using the photometric membership criteria, the foreground and background galaxies are removed (see Rakos et al. 1991 for a full description of the procedure) such that the member galaxies are found to be within 1000 km s$^{-1}$ of the cluster mean redshift. This membership criterion resulted in the cluster \objectname{A779} having 45 galaxies in our 1$^{\circ}$ field of view. These member galaxies range from $13.3 \leq m_{5500} \leq 20.0$. Spectroscopic studies, on the other hand, of Hwang \& Lee (2008) using 3$^{\circ}$ field of view of SDSS and 2$^{\circ}$ field of view of 2dFGRS find 145 cluster members for A779.

\section{Results}

\subsection{\textit{Method for Age and Metallicity Estimation}}

In the last two decades, the choice of the techniques for estimating the ages and metallicities of galaxies has been through the following. (1) Spectral Energy Diagram (SED) fitting of the age and metallicity sensitive spectral indices, through Lick indices (Trager et al. 2000, Gonz\'alez 1993). (2) Spectrophotometric indices, for measuring the strength of specific features that are not sensitive to the flux calibration (Koleva et al. 2008). (3) Full spectrum fitting, that uses all the redundant information contained in the signal. This method is a combination of spectrophotometric indices and SED fitting (Koleva et al. 2008). Nevertheless, various age sensitive indices are now found to contaminate the metallicity sensitive indices and vice versa, that result in inaccurate and/or incorrect estimations of the two parameters (Worthey 1999). Ferreras et al. (1999) point out the age sensitive H$\beta$ line is significantly affected by the nuclear emissions in the E/S0s for which corrections are uncertain (Gonz\'alez 1993) and the [Mg/Fe] index changes drastically from the faint to the bright E/S0s, which challenges their use as a metallicity index.

A wealth of information on various age or metallicity sensitive spectral lines confuses their estimation. The strength of the Fe index hardly shows any dependence on the galaxy mass (Fischer et al. 1996). Sanchez-Blazquez et al. (2006) found that by using $Mg_2$ index, no correlation in age and metallicity can be made; instead a clear trend is observed with the Ca and Fe indices. These challenges result in difficulties in painting a proper picture of galaxies. Since the evolutionary models do not use individual line strengths, but the combined strength of several line indices which only point out the star's position on the Hertzsprung$-$Russel (H$-$R) diagram (Rakos et al. 2007). These ages, determined using the spectral indices, have recently been contradicted by many methods, for example, the Lick/IDS system which find many cluster ellipticals of relatively younger ages of less than 6 Gyr (Sanchez-Blazquez et al. 2006), the photometric system using continuum colors detect only a very few ellipticals less than 10 Gyr.

On the other hand, our method has the advantage of measuring the dominant component that produces the observed galaxy colors, which is the continuum emission (Rakos et al. 2007) with high signal-to-noise ratio at relatively short exposures. The continuum colors are the primary source of information related to the characteristics of the underlying stellar populations. The age and metallicity determine the continuum emission through the impact of the higher mass stars in the younger stellar populations (hot stars) and the changes in the color of the RGB by line blanketing effects caused by the variation in the mean metallicity (Tinsley 1980), respectively.

R05a applied the PCA\footnote{The program could be downloaded at \textit{http://student.fizika.org/$\sim$vkolbas/pca/index.html}} (PCA) method (Steindling et al. 2001) to the S02's Single Stellar Population (SSP) models that are based on resolved isochrones of the Padova group. Briefly, the PCA in any $n$-dimensional space calculates the axis with the most significant scatter as the PC1, then the second most significant scatter in the remaining $n-1$ dimensional space, orthogonal to PC1 as PC2 and so on. The advantages of applying the PCA are to reduce the dimensionality of the data and to provide the orthonormal coordinate system in which the data are most easily characterized. The original PC equations that are only color dependent are:

\begin{eqnarray}
PC1 = 0.80(uz-vz) + 0.53(vz-bz) + 0.28(bz-yz)
\end{eqnarray}
\begin{eqnarray}
PC2 = - 0.56(uz-vz) + 0.49(vz-bz) + 0.67(bz-yz)
\end{eqnarray}
\begin{eqnarray}
PC3 = 0.22(uz-vz) - 0.69(vz-bz) + 0.69(bz-yz).
\end{eqnarray}

The PCA is an empirical method based on the catalogs of galaxies and stellar population models (Kennicutt 1992a, 1992b; Kinney et al. 1996), which when applied to the S02 models offers the following advantages. (1)  The PCA method helps in the selection of the cluster members using "the cluster box" (refer Figure 2 in Steindling et al. 2001). (2) The classification of galaxies into different Hubble types is possible using this analysis technique. The PCA allocates specific places for different galaxy morphological classes in "the cluster box" built by the three PC equations ((1)$-$(3)). (3) Importantly, the PCA separates the age effects from the metallicity effects.

In order to apply the PCA, the required linearity was observed for the SSPs with ages older than 3 Gyr for a full range of model metallicities. The primary reasons for the application of the PCA are to separate the age and metallicity of a stellar population, select the most correlated variables, and to determine the linear combination of variables for extrapolation. Therefore, R05a adjusted these PC equations to include the model age and metallicity terms along with the observed colors as shown below

\begin{eqnarray}
PC1 = 0.471[Fe/H] + 0.175\tau + 0.480(uz-vz) + \nonumber \\
0.506(bz-yz)  + 0.511(vz-yz)
\end{eqnarray}
\begin{eqnarray}
PC2 = 0.345[Fe/H] - 0.935\tau + 0.047(uz-vz) - \nonumber \\
 0.061(bz-yz)  + 0.020(vz-yz),
\end{eqnarray}

where $\tau$ is the age of the stellar population in Gyr. Metallicity terms in both the equations have almost equal weights, but the weights for the age terms are smaller in the first equation compared to the second. The SSP models reflect the ages and metallicities of star clusters and the summation of the star clusters would reflect the age and metallicity of the composite stellar populations, like, the ellipticals. However, the low-order difference between globulars and ellipticals signals that the SSP assumption is an adequate approximation for many needs and that the stellar population in ellipticals must be of a simple form, like, a Gaussian distribution to produce correlation (R05a).  Since the filter system is more sensitive to the metallicity than to the age, the uncertainty due to the age is greater than for the metallicity which is 0.5 Gyr and 0.2 dex, respectively, as quoted by R05a.

The other model considered here for the comparison is the GALEV (successor of the S02 models; Kotulla et al. 2009), a galaxy evolutionary synthesis model that produces a large grid of galaxy spectral energy distribution templates in a chemically consistent manner. The GALEV offers only five metallicity models: $-$1.7 dex, $-$0.7 dex, $-$0.3 dex, 0.0 dex, and 0.3 dex. For more details, refer to Kotulla et al. (2009).

\subsection{Comparison of Colors, Model Ages, and Metallicities}

This section discusses the results for the following tests to check. (1) The repeatability of the observed narrowband colors using the two filter systems, the Str\"{o}mgren and the modified Str\"{o}mgren systems, to assess the quality of the observed colors.  (2) The agreement of the estimated ages and metallicities using two different techniques: the PCA and the GALEV models.

Unfortunately, from the available data, only six bright, isolated galaxies were found common in all the three data sets. These galaxies are in the magnitude range of $14.1 \leq m_{5500} \leq 15.5$ ($-20.7 \leq M_{5500} \leq -19.3$). Figure 1 shows the combined \textit{bz} (4675 \AA) image of \objectname{A779} observed at Steward 90 Prime in 2004 April with the chosen galaxies numbered 1$-$6. Table 2 gives the comparison of colors, $M_{5500}$ brightness between the three sets of observations along with their age and metallicity estimations using the PCA and the GALEV models. The mean displayed for each set of galaxy colors, ages and metallicities represents their close nature, and hence smaller uncertainties.

Table 3 presents, object wise, the standard deviations ($\sigma$) in the measurements of the narrowband colors and their respectively derived ages and metallicities using the two evolutionary models. A low standard deviation conveys the quality of repeatability of the observed colors. The \textit{(uz-vz)} color has the largest standard deviation ($\sigma$) value of 0.09 amongst the three, because of its sensitivity to the atmosphere, whereas the \textit{(bz-yz)} and \textit{(vz-yz)} colors have 0.02 and 0.03, which are small and expected due to photometric errors and the  atmospheric extinction. The mean $\sigma$ value for the ages and metallicities determined by the PCA method are found to be 0.36 Gyr and 0.04 dex, respectively, within the uncertainty limits (0.5 Gyr in age and 0.2 dex in metallicity) as quoted by R05a. On the other hand, the GALEV estimates the same two parameters with the accuracy of 0.44 Gyr and 0.2 dex for the age and metallicity, respectively. The mean of the standard deviation of the GALEV estimated ages and metallicities are found to be larger compared to the PCA, which are due to GALEV's limited range of model metallicities ($-$1.7, $-$0.7, $-$0.4, 0.0, and 0.4 dex) and a significant effect of the age$-$metallicity degeneracy, that could be relieved with the use of high-resolution spectral library and the application of the PCA. Nevertheless, the low values of $\sigma$ indicate a fair similarity between the two model age and metallicity estimations.

The observed differences in colors, ages, and metallicities could be explained by the following reasons.  (1) The uncertainties in the colors (of the order of a few hundredths of a magnitude) could be due to the atmospheric extinction and photometric errors. (2) The uncertainties of 0.5 Gyr in age and 0.2 dex in metallicity can result from the interpolation between the limited range of the S02 models, and the iterative method used by the PCA technique. (3) The ability of the individual models, and the limited range of the model metallicities of the S02 and the GALEV, to reproduce the correct ages and metallicities from the observed galaxy colors can also induce errors. (4) The effect of prevailing age$-$metallicity degeneracy can further affect the results especially in the GALEV models, as it is not supported by any additional technique to separate the age effects from those of metallicity.

The effect of the age$-$metallicity degeneracy has been a major obstacle in all the stellar population evolutionary techniques in their accurate determination. The continuum color method supported by the PCA technique does, in fact, separate the age effects from the metallicity effects, but only to a certain extent (as shown by the standard deviation in Table 3). From Table 1, in R05a, close similarities in the specific colors for different age and metallicity SSPs  can be found, for example, the \textit{(uz-vz), (bz-yz), (vz-yz)} colors for the $Z$ = $-$0.7 dex, age = 6.02 Gyr are 0.713, 0.232, 0.360 and for $Z$ = $-$0.4 dex, age = 4.06 Gyr with 0.718, 0.231, 0.375, respectively, are quite similar. Likewise, the SSP with $Z$ = $-$0.4 dex, age = 6.02 Gyr, the \textit{(uz-vz), (bz-yz), (vz-yz)} colors are 0.731, 0.256, 0.443 which are similar to $Z$ = $-$0.7, age = 8.12 Gyr (with colors 0.727, 0.258, 0.432) and $Z$ = $-$0.7, age = 10.08 Gyr (0.731, 0.260, 0.442).

Observationally, these color differences are well within the photometric uncertainty limits, which may result in incorrect estimations of the ages and metallicities of the SSPs. But when considering a galaxy with the luminosity-weighted mean age and metallicity, these uncertainties are masked under the galaxy's overall age and metallicity. Obviously, this would depend upon the mass of the SSP (of a certain age and metallicity) that is the major contributor to the galaxy's overall age and metallicity. For example, in an old red galaxy which is a mixture of several SSPs (mostly old), it is very difficult to estimate the uncertainty of the SSP with incorrect age and metallicity, because of the indistinguishability of the masses of different SSPs which constitute to form a single galaxy age and metallicity (Rakos et al. 2008). But for a young galaxy with most of its star clusters primarily young with perhaps similar ages and metallicities, the uncertainty in such a case would be comparatively higher. However, this error and its effect is more pronounced in studies of individual star clusters and young galaxies.

Nevertheless, our analysis of different SSP ages and metallicities conveys that the degeneracy is still prevailing. In order to minimize the effect of this degeneracy, it is important to reduce the photometric uncertainties by avoiding very faint and diffused (galaxies with large gas content) objects and choosing judicious apertures depending upon the brightness and the type of galaxy, as performed in our studies. Since our photometric studies are based on Str\"{o}mgren (or modified Str\"{o}mgren filters) that are well known for their age, metallicity sensitivity and are supported by the PCA technique, the effect of degeneracy observed here is not as severe as reported by spectroscopic and broadband observations. Therefore, the degeneracy is only relieved, though not broken.

\section{Discussion}
\subsection{\textit{Spectrophotometric Classification}}

Using the PCA classification for galaxies, based on the narrowband colors, our sample segregates into four main categories, Ellipticals (E, passive, red objects), S$-$ (transition objects between ellipticals and spirals) types like the S0s, Spirals (S, objects with their star formation rate equivalent to the disk galaxies), and S+ (starburst systems). By comparing nearby galaxy observations with the SED models, the divisions along the PC1 axis are made such that the S galaxies are the systems with Spiral star formation rate of about 1 $M_{\odot}$ yr$^{-1}$ and the S+ objects correspond to starburst rates of about 10 $M_{\odot}$ yr$^{-1}$ (Rakos et al. 1996). Once again, one must remember that these star formation rates represent the mean over the last few gigayears, and not the current star formation rate, as reflected by the optical emissions of the dominant stellar population. The colors displayed by the red E systems show no evidence of star formation in the past 4$-$5 Gyr. The S$-$ class of objects is the transition galaxies that show no sharp division between the E and S galaxies. These galaxies show slightly bluer colors (statistically) from the passive E galaxies; however, the difference could be due to a recent, low-level burst of star formation or later epoch of galaxy formation or an extended phase of early star formation or even lower mean metallicity (i.e., the color$-$magnitude effect). The PC2 is able to classify signatures of non-thermal continuum colors of the active galactic nuclei intoÊ A+, A, A- types (for more descriptions, see Steindling et al. 2001 and Rakos et al. 2007).

The PCA classification scheme shows that our \objectname{A779} sample constitutes one-third (31\%) of the ellipticals, 40\% of the S$-$ types (S0s) forms the majority,  the S class are 24\% and the S+ are just 4\%. This classification scheme, based on the PC equations using the narrowband color indices, agrees with the results of Hwang \& Lee (2008), in finding more numbers of the early type (ellipticals and S0s) galaxies than the late types. Table 4 lists the A779 member galaxies with their names, R.A., declination and their photometric errors. Since Rakos et al. (2008) have previously studied and analyzed the color$-$color and color$-$magnitude relations for the galaxies in A779, we refrain from repeating the same inference in this paper. For a more detailed color$-$color and color$-$magnitude relations, see Rakos et al. (2008).

\subsection{\textit{Spectrophotometric Class$-$Density Relation}}

Even though, the continuum colors are integrated in the PC equations (needed for galaxy classification), a majority of the galaxies classified using this technique are found to well match the morphological classifications. The idea is to build a close one-to-one relationship between the spectrophotometric classification and the morphology (Rakos \& Schombert 2005b). Figure 2 shows the variation in the density of spectrophotometric galaxy classes as a function of radial distance from the \objectname{A779} cluster core. The galaxy density as a function of spectrophotometric class is binned in an area of 0.1 Mpc$^2$ extending out from the cluster center, resembling the density$-$morphology relation (Dressler 1984; Smith et al. 1997). Due to the small sample of the S+ objects (and the similarities in their star formation rates), we have combined them with the S class of objects. The figure also displays large errors for different spectrophotometric class due to the fewer A779 galaxies that are spread sparsely across the cluster distance.

The figure shows the innermost regions of \objectname{A779} which are dominated by the red population, largely by the S$-$ and some E types, in agreement with Wilson \& Vall\'ee (1982) and Fanti et al. (1984).Ê The star-forming blue populations, the S/S+ types, avoid cluster cores and are found to populate at 0.2 Mpc from the center, and their density declines with increasing distance. Most of the passive galaxies are known to occupy regions very close the cluster core, but, surprisingly, the E and the S$-$ types also show their presence at distances of 0.3 Mpc from the center. These galaxies could represent the second generation of young red population with a small mass of star-forming populations.

Another inference that can be gathered from Figure 2 is with respect to the changing ratio of the blue to red cluster galaxies, the so-called Butcher$-$Oemler effect (Butcher \& Oemler 1984). Blue galaxies, those concentrated at 0.2 Mpc from the cluster center are slowly turning into red, like the spirals to the Sa types (represented by the S$-$ types occupying the region near 0.3 Mpc from the cluster center). These S$-$ galaxies have low star formation rates, and as they age their star formation ceases and they drift to the cluster centric regions, joining the S0 populations at 0.1 Mpc. The E class of red galaxies cannot be the rightful candidates for the blue cluster galaxies because the majority ofÊtheir mass consists of old (12 Gyr) populations.

\subsection{\textit{Ages and metallicities of A779 cluster galaxies}}

The optical colors of galaxies reflect information that pertains to the photospheres of the stars present in stellar populations, and any change in these colors would indicate the change in these stellar populations. Since stellar populations keep a record of most of their evolution, by studying the H$-$R diagram (or color$-$magnitude diagrams) of these stars, it is possible to constrain the star formation history (age) and enrichment (metallicity) from their initial formation (Grebel 2005). After testing for the precision of the estimated ages and metallicities using the PCA and the GALEV models, we present here the luminosity-weighted SSP mean ages and metallicities for \objectname{A779} member galaxies using the PCA method. These ages and metallicities are shown in Figure 3 along with their correlation with $M_{5500}$ luminosity (mass). The typical uncertainties for the ages, metallicities, and $M_{5500}$ are 0.5 Gyr, 0.2 dex (R05a), and 0.15 mag, respectively, displayed at the corner of each plot. The differing sizes of the data points in the plots show the true dispersions in the averages of the SSP ages and metallicities, and are not uncertainties. Since these true dispersions depict the standard deviation of the averaged SSP ages and metallicities, it would represent the star formation histories, in such a way that higher the dispersion due to longer star formation history, bigger would be the size of the data point. 

Directly visible from the plot in Figure 3 is the "bimodality" of two clumps of populations divided by their age, metallicity, and mass. The first group of objects is the older mixture of the early-type ellipticals and the S0s, with the mean age of 9 Gyr, classified as the classic red-sequence galaxies. These red-sequence galaxies show a horizontal spread in metallicity ranging from $-$0.5 dex to slightly above solar. The second group is the blue-sequence galaxies, that show a mixture of the S$-$ class, spirals (S), and the starburst (S+) objects, which are younger (5 $-$ 8 Gyr) than the red sequence galaxies. Also, compared to the older galaxies, these younger galaxies show relatively stronger chemical evolution with $Z$ ranging from $-$2.1 dex to slightly above $-$0.5 dex.

Interestingly, the galaxy quantities, like the age, $M_{5500}$ (mass), and metallicity, are found to correlate well with one another; the low-mass galaxies are metal-poor ($Z$ $\leq$ $-$0.5 dex), whereas the metal-rich ($\sim$ $Z_{\odot}$) galaxies are more massive and older galaxies. Galaxies with ages of 8 Gyr are brighter than $M_{5500}$ = $-$18 mag with near solar metallicities, whereas the young, low-mass galaxies are metal-poor with $Z$ $<$ $-$0.5 dex. This phenomenon of decreasing mass with age is referred as the "Downsizing" scenario of galaxy formation (Bundy et al. 2005; Mateus et al. 2006). The best demonstration of the downsizing effect is the color$-$magnitude relation diagram from $z$=0.7 clusters by De Lucia et al. (2004), that show deficiency of low-luminosity red galaxies between $M_{v}$= $-$19 mag and $-$17 mag. Studies from Rakos et al. (2007) show that only 5\% $-$ 10\% of the galaxy's mass needs to be involved in star formation to produce the spiral-like colors, as seen in the De Lucia et al.'s diagrams. Given that $z$=0.7 corresponds to 6 Gyr ago, this would produce the current generation of low-luminosity cluster ellipticals with 10\% of their galaxy mass in a 6 Gyr population plus the remaining 90\% of their galaxy mass in a 12 Gyr population (Rakos et al. 2007). Convolving these values in our SED models reveals that the luminosity-weighted mean age of such a galaxy would be about 10 Gyr, as observed in the plots of Figure 3. The blue population (S$-$, S, and S+ class) consists of the star-forming galaxies which slowly fade to the low-mass S$-$ class of the S0s with traces of star formation. The blue colors of this galaxy type show the ongoing star formation with a majority of their stellar mass in the old stellar populations and a substrate of young (less than 1 Gyr) stellar population that combine to produce an integrated age of 5$-$7 Gyr. Once again, using the SED models we find, that about 10\% of the mass of these blue galaxies can be in a 1 Gyr stellar population and the rest in a 12 Gyr one to produce the observed colors of about 6 Gyr. 

Thomas et al. (2002) claim that [$\alpha$/Fe] ratios are highest in the oldest galaxies of his sample, which would imply a short initial duration of star formation for the old galaxies. This also conveys that the initial star formation duration increases with decreasing mass, as evident from our results that show, on average, the large sizes of the data points for the young, blue-sequence galaxies compared to the old population. A set of blue asterisk symbols in Figure 3 represent the S (spiral) population; these are low-mass, young galaxies with metallicities between $-$0.5 dex $\leq$ $Z$ $\leq$ $-$1 dex and age range of 5 Gyr $\leq$ $\tau$ $\leq$ 8 Gyr. A careful inspection of these S galaxies with their extended star formation histories and their location from the cluster confirms the 21 cm radio study of Solanes et al. (2001), that the inner and outer 1 Abell radius ($\sim$ 1Mpc) of this cluster have their spirals with significantly low HI content. This could be due to the star formation events that transformed the HI gas to H$_2$ to form stars in these low-mass spirals localized at 0.3 Mpc from the cluster center. 

The presence of only two starburst galaxies in our sample, one very close to the cluster core and other at the outer regions of the cluster, is somewhat strange. As starbursts represent a collection of young objects (OB associations of a few million years old), their analysis is not quite reliable, as our technique is only effective for stellar populations with ages greater 3 Gyr. The luminosity-weighted mean ages of the two starbursts in our sample are found to be 6.4 Gyr and 7.8 Gyr, at 0.06 Mpc and 0.3 Mpc from the cluster center, respectively. The presence of a starburst galaxy near the cluster core could be ruled out as a projectional effect, since it is highly unlikely for a starburst system to be actively forming stars at such close proximity to the cluster core. However, the other at 0.3 Mpc from our morphological analysis performed on our images shows similarities to an irregular disk system with high star formation history. While dealing with systems that have high star formation rates and their bluer colors, it is possible that our spectrophotometric classification scheme miscategorizes some high star-forming spirals/irregular disk systems as starbursts, because of the close resemblance of the galaxies in the S and the S+ category.

\subsection{\textit{Spectrophotometric Class-Age Histogram}}

Figure 4 shows a normalized histogram of age for different spectrophotometric classes (in steps of 1 Gyr above 5 Gyr) of \objectname{A779} galaxies. In agreement with the above discussion, this figure, in general (ignoring the galaxy classification), presents the bimodality of old population that peaks at 8 Gyr and the young population at 6 Gyr. Also, visible in this histogram, within the two groups there are differences in age for different spectrophotometric types. But when observed as a function of spectrophotometric types, the bimodality feature is observed only for the S$-$ types (S0s), as the S0s forms the majority in our cluster sample. The two groups of S0s show peaks for young 6 Gyr (about 8 in number) galaxies, which matches the peak of the maximum number of the star-forming S/S+ types, and another for the 8 Gyr old S0 galaxies, that matches with the maximum of the E galaxies. This clearly shows that the S0s are the transitional galaxy types that bridge the gap between the red- and blue-sequence galaxies. The old (8 Gyr) E galaxies are mostly non-star-forming bulge systems, whereas the second group of young galaxies consists of the S$-$, S, and S+ galaxies that show a mixture of old, bulge system plus a younger disk population.

With regard to the star formation history, the second group of the S$-$ galaxies does show the possibility that a small percentage of their mass are star forming, which in result produce their younger mean ages. The determination of the exact nature of these young S$-$ classes is difficult because our technique is unable to distinguish, using SED models, between star formation history scenarios where the young population is composed of a young age due to (1) later formation redshift, (2) longer initial star formation duration, or (3) a later burst of star formation (a ÔÔfrosting eventÕÕ; Trager et al. 2000). 

\subsection{\textit{Age/Metallicity$-$Distance Relation}}

Figure 5 presents the variation in the age and metallicity of the cluster galaxies as a function of cluster centric distance for different spectrophotometric classes and their star formation history. Due to lack of sufficient sky coverage, we are restricted to analyze galaxies within 0.4 Mpc distance from cluster center. Unlike in the evolved clusters (e.g., A1185, Coma, A2218), the plot shows that \objectname{A779} shows no significant trend between the galaxy age with distance from the cluster center and the old early types are distributed randomly across the cluster. The E galaxies near the cluster core are about 9 Gyr, which vary only slightly to 8 Gyr in age up to a distance of 0.25 Mpc, and then rise to 10 Gyr at 0.3 Mpc. A similar trend is also observed for the S$-$ types with a 2 Gyr dip in age for galaxies near the core to a distance of 0.25 Mpc and again a 2 Gyr rise for the galaxies at a distance of 0.35 Mpc. For the late types, the variation in age with distance is again very small, from 6 Gyr at 0.1 Mpc to about 7 Gyr at 0.4 Mpc.

As the galaxy age and metallicity behave in the same direction as observed in Section 4.3, it is not surprising to expect the trend of metallicity-distance to be very similar to the age$-$distance relation. But compared to the haphazard pattern of the age$-$distance plot, the metallicity shows slightly organized separation for the different spectrophotometric class with distance, such that, the E galaxies have near solar metallicities, the S$-$ types of about $-$0.4 dex and the S/S+ objects with $-$0.8 dex. Since different galaxy types evolve differently under the influence of differing environmental effects across the cluster radial distance, we analyzed for the differential variation in metallicity for 1 Gyr change in age for the three spectrophotometric classes across the radial distance. We note a significant change in metallicity for the E types at 0.1$-$0.2 Mpc, the maximum variation in metallicity for the S$-$ class is observed at 0.2$-$0.3 Mpc and 0.25$-$0.4 Mpc for the S/S+. Interestingly, the region of 0.2$-$0.4 Mpc from the cluster core is the most dense with most (22) of our galaxy samples, which, on average, show large star formation histories. Clearly, considering the direction toward the increasing differential variation in metallicity and the presence of a large number of galaxies with high star formation histories, the region beyond 0.25 Mpc seems to be the breeding ground with a lot of active physical processes to enhance the star formation in the young low-massÊ galaxies (S types with ages between 5-7 Gyr and the second generation of the S$-$ types of 5$-$9 Gyr), as also confirmed by our Figure 2. As a further confirmation, these low-mass, young star-forming galaxies present a scattered population in Figure 3, which distinctly show the physical processes active in the galaxies located in these regions. The presence of two large E data points in this region was morphological analyzed to find their resemblance to the E5$-$E7 elliptical galaxy types that are known to be misclassified S0s (van den Bergh 2009). 

As an overall analysis ofÊ \objectname{A779} using the modified Str\"omgren photometry, we find that the cluster is fairly young, low mass and has a poor dynamical evolution which is reflected by its fewer members, disorganized distribution of galaxies by age and metallicity across the radial cluster distance and the low star formation histories of the member galaxies compared to other dynamically evolved clusters.

\section{Conclusion}

In the first part of this paper, we tested six galaxies in \objectname{A779} for the repeatability of colors using the three sets of observations with the two filter systems: Str\"omgren and modified Str\"omgren. Our test shows a fair repeatability that reproduces similar ages and metallicities within an uncertainty of 0.5 Gyr in age and 0.2 dex in metallicity, as presented in Tables 2 and 3. The ages and metallicities are estimated using the PCA method and the GALEV models. The mean standard deviation $\sigma$ for colors \textit{(uz-vz)}=0.09, \textit{(bz-yz)}=0.02, and \textit{(vz-yz)}=0.03 are found. The $\sigma$ for the measured ages and metallicities from the PCA method are found to be 0.04 Gyr and 0.36 dex, respectively and from GALEV 0.2 Gyr and 0.44 dex, respectively.

The small observed color differences could be accounted: (1) from the photometric errors,  (2) slight variation in the two filter transmission curves, (3) $\lambda$ dependent flux measurement of the modified Str\"{o}mgren system. The uncertainties in the estimated ages and metallicities could be due to (1) the quality of the solutions obtained by the PCA iterative method, (2) the ability to accurately estimate ages and metallicities by the S02 and the GALEV models, (3) their limited model metallicity range to reproduce proper ages and metallicities, (4) the effect of the age-metallicity degeneracy problem in the GALEV models, and (5) the errors due to interpolation for the S02 models using the PCA method.

Careful inspection of the SSP colors for different ages and metallicities enlisted in Table 1 of R05a shows close similarities between one another within the limits of photometric uncertainties. This conveys that even with the application of PCA, to separate the age effects from the metallicity effects, the degeneracy still persists, but the effect is not as significant as observed from the spectroscopic and broadband observational studies. Therefore, the age$-$metallicity degeneracy using narrowband continuum color method is only relieved, though not completely broken.

In our second part, we analyze the X-ray and radio quiet, poor \objectname{A779} cluster with 45 member galaxies detected by our photometric method in the 1$^{\circ}$ field of view. We note a clear correlation of galaxy age, metallicity, and luminosity (mass) to understand their formation scenarios with effect to the cluster environment. Using our spectrophotometric classification method, we find that the high-mass, luminous (E/S$-$) galaxies have a luminosity-weighted mean age of 9 Gyr ($\pm$ 0.5 Gyr) and the low-mass, fainter (S/S+ and the young E/S$-$) galaxies of the mean age of 6.7 Gyr ($\pm$ 0.5 Gyr). The older, red-sequence galaxies show higher mean metallicities of $-$0.18 dex ($\pm$ 0.2 dex), whereas their younger, blue-sequence counterparts are slightly metal-poor with the mean of $-$0.63 dex ($\pm$ 0.2 dex). The galaxies with the higher mean age and metallicity are more massive than the younger, low mean metallicity objects. The older galaxies comprise mostly ellipticals (E class) and the S0s (S$-$ class), whereas the young galaxies are a mixture of the spirals, disks/irregulars (S class), the starburst systems (S+), and the low-mass red sequence of galaxies.

We observe the usual trend of galaxy age with mass, such that the older, red population have a single short epoch of star formation and this initial duration of star formation increases with decreasing mass, a weaker version of the downsizing effect (Gallazzi et al. 2006). A similar phenomenon is also observed by the smaller spread in metallicity for the red sequence galaxies, hinting a common origin for these populations, and the higher luminosities point to their massive nature. The scatter in the second group (blue sequence) of galaxies points to the ongoing physical processes that show an extended chemical evolution with the longer epoch of star formation. The star formation histories of the second group of galaxies also shows the presence of a few low-mass S$-$ galaxies which have a small percentage (5\%$-$10\%) of young, low-mass star-forming population inside the old S0 galaxies.

The cluster's innermost regions are occupied by the old, red-sequence galaxies, whereas the blue star-forming galaxies are present at 0.2 Mpc from the cluster core. These low-mass S/S+ galaxies represent the Butcher$-$Oemler galaxies that would fade to the red Sa types to join the other second generation of the S$-$ types at 0.3 Mpc from the cluster center. As the traces of star formation ceases with age, these S$-$ galaxies would drift to settle with the other S0s observed close to the cluster center at 0.1 Mpc.

The correlation between the age (and metallicity) with radial distance from the center, as a function of spectrophotometric class and the star formation history, is poor. The old early-type galaxies follow a random distribution by age and metallicity across the cluster distance from the center. Though, the late-type, young galaxies do show a slight trend of 6 Gyr galaxies at 0.1 Mpc to about 7 Gyr at 0.4 Mpc. The metallicity$-$distance trend is similar to that of the age$-$distance relation. The plot confirms our idea about the blue galaxies which are actively forming stars at 0.25 Mpc, slowly drifting to the cluster edges as they age and eventually fall into the S$-$ types at the cluster center. The region beyond the 0.25 Mpc appears to be the breeding ground for low-mass galaxies with a lot of active physical processes leading to a larger chemical evolution and enhanced star formation.

Overall assessment, which is in agreement with other studies using different wavelengths and methods, indicates the cluster \objectname{A779} as a young, low-mass, dynamically poor cluster with a passive cD galaxy (NGC 2832) close to the cluster center. Our technique of the modified Str\"omgren photometry has also shown an in-depth analysis of the cluster galaxies with the age and metallicity measurements and their correlation with spectrophotometric galaxy types, density, star formation history, and cluster's radial distance. This analysis proves the technique's ability to study galaxy clusters at low- and intermediate-$z$.

\section{Acknowledgments}

We are grateful to our colleague and mentor, Karl Rakos (86), passed away on {2011 October 31}, and we would like to dedicate this paper to him and his contributions to the field. We thank the team at Steward Observatory, University of Arizona for the time allocation and the assistance and support that they provided. This work is supported and funded by the University of Vienna within the Initiative College "Cosmic Matter Circuit" (IK 538001).

\clearpage

%% the file, uncomment the \plotone, \plottwo, and
%% If you need a layout that cannot be achieved with \plotone or
%% \plottwo, you can invoke the graphicx package directly with the
%% \includegraphics command or use \plotfiddle. For more information,
%% please see the tutorial on "Using Electronic Art with AASTeX" in the
%% documentation section at the AASTeX Web site,
%% http://www.journals.uchicago.edu/AAS/AASTeX.
%%
%% The examples below also include sample markup for submission of
%% supplemental electronic materials. As always, be sure to check
%% the instructions to authors for the journal you are submitting to
%% for specific submissions guidelines as they vary from

%% This example uses \plotone to include an EPS file scaled to
%% 80% of its natural size with \epsscale. Its caption
%% has been written to indicate that additional figure parts will be
%% available in the electronic journal.

%\begin{figure}
%\epsscale{.80}
%\plotone{A779bimage.eps}
%\caption{Abell 779 cluster at z= 0.023 with a central cD galaxy (NGC 2832), stretched N-S direction. This $0.3\arcmin$ x $0.3\arcmin$ combined image of three \textit{bz} ($4675\AA$) filter image was observed on 2004 at Steward Observatory, Arizona. The figure shows the six common galaxies in all the three datasets which are compared in Table 2 and 3.{\itThe Astrophysical Journal}.\label{fig1}}
%\end{figure}

\clearpage

%% Here we use \plottwo to present two versions of the same figure,
%% one in black and white for print the other in RGB color
%% for online presentation. Note that the caption indicates
%% that a color version of the figure will be available online.
%%

\begin{figure}
\includegraphics[angle=0,scale=0.8]{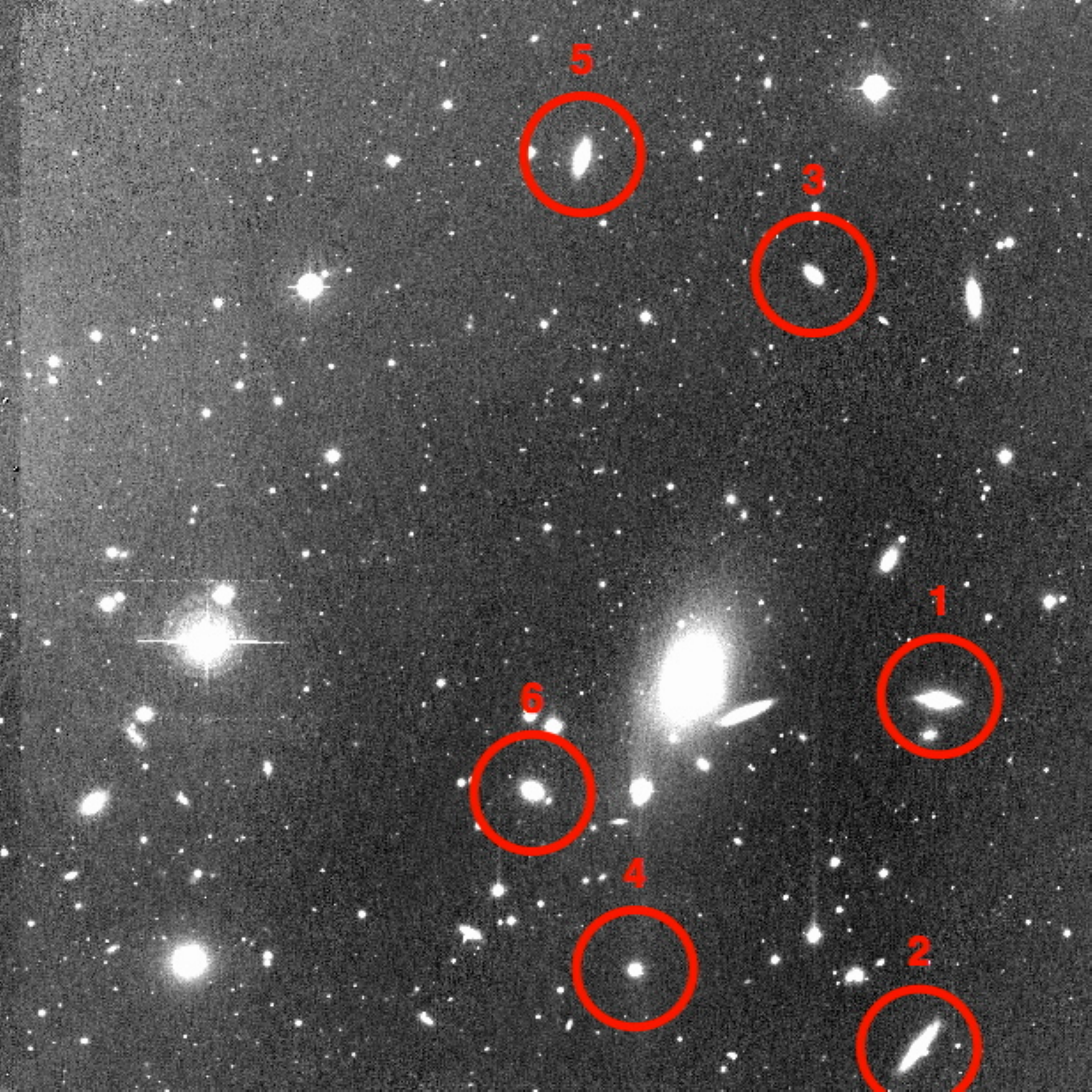}
\caption{A 779 cluster at $z$= 0.023 with a central cD galaxy (NGC 2832), stretched N$-$S direction. This $0\arcmin.3$ $\times$ $0\arcmin.3$ combined image of three \textit{bz} ($4675$\AA) filter image was observed on 2004 at Steward Observatory, Arizona. The figure shows the six common galaxies in all the three data sets which are compared in Tables 2 and 3.}
\end{figure}

\begin{figure}
\centering
\includegraphics[scale=1.2]{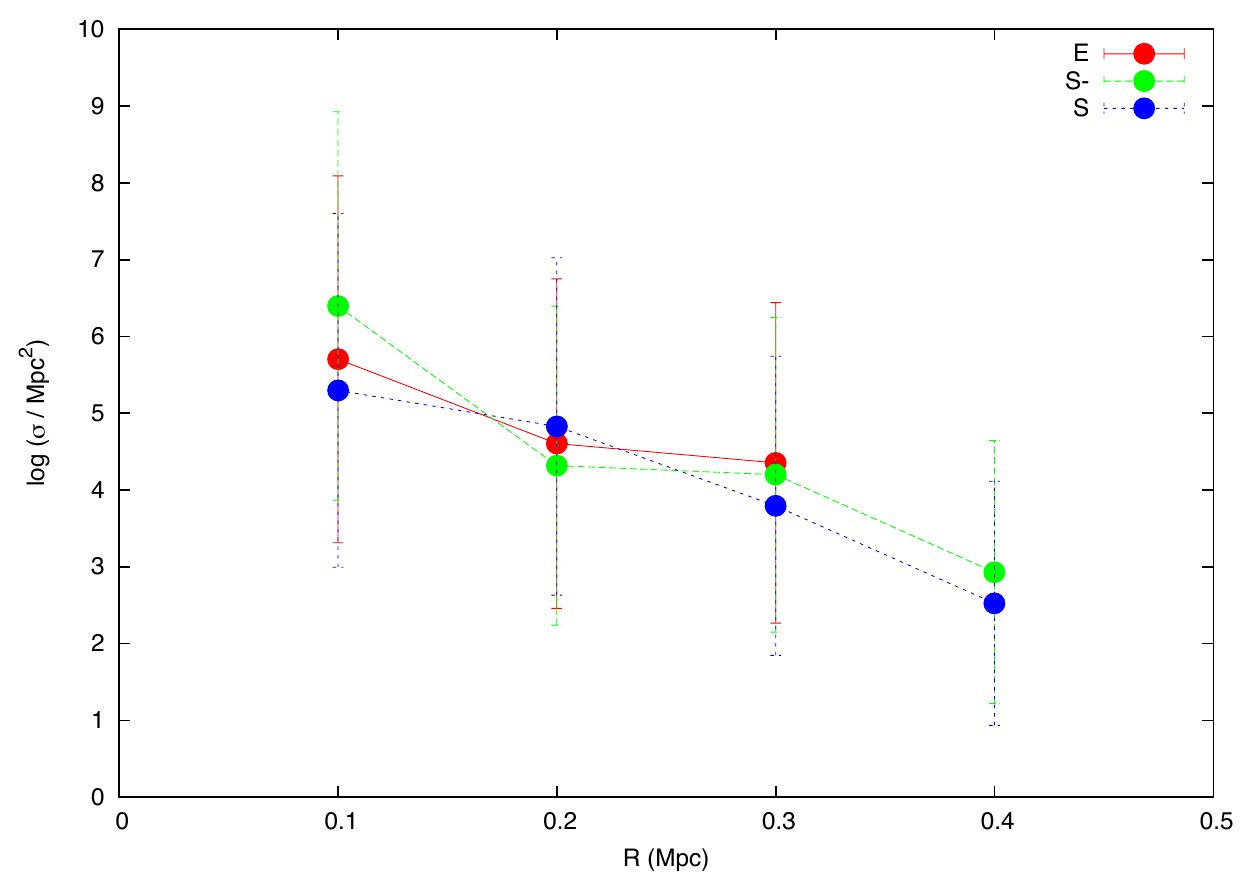}
\caption{Projected radial galaxy density for each galaxy spectrophotometric class as a function of radial distance from the cluster center. The number of spectrophotometric galaxy type binned in an area of 0.1 Mpc$^2$ from the cluster center.The plot displays the central cluster regions are occupied by the old S$-$ galaxies, that fall almost sharply with a slight increase in the blue S/S+ star-forming galaxies at about 0.2 Mpc from the center. At 0.3 Mpc, there is a slight increase in second generation of the young S$-$ and some E galaxies.}
\end{figure}

\begin{figure*}
\centering
\includegraphics[scale=0.65]{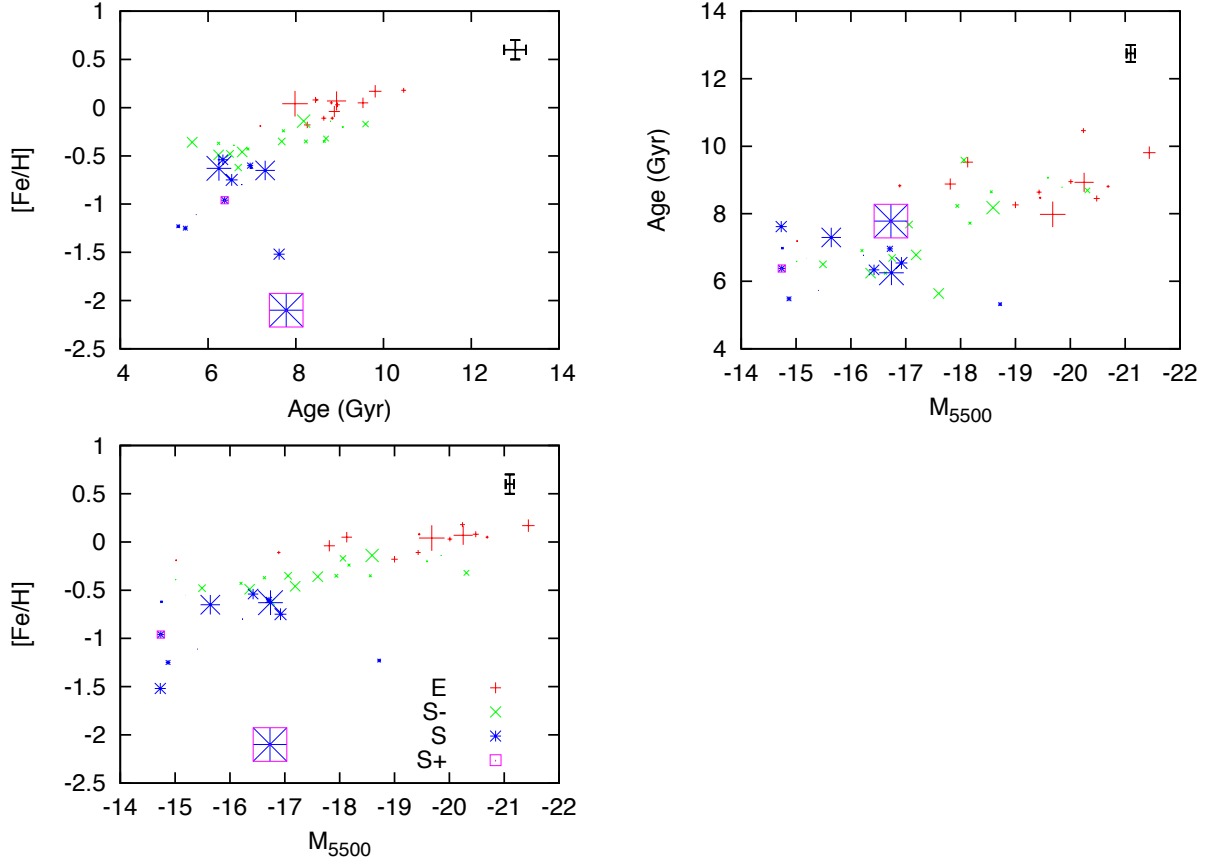}
\caption{Age, $Z$, mass (luminosity) plot for A779 galaxy members. Each plot compares the SSP luminosity-weighted mean, age, and metallicities, for all 45 A779 member galaxies. Different galaxy classes are displayed with different colors (E types in red pluses, S$-$ in green crosses and S in blue asterisks and the S+ in purple squares). The typical uncertainties for age are 0.5 Gyr, $M_{5500}$ are 0.15 mag, and the $Z$ are 0.2 dex, which are shown at the corner of individual plot. The varying size of individual data points gives true dispersion in the averages. Two distinct groups of populations are observed, one with the mean age of 9 Gyr, represented by the old, massive, metal-rich, red populations, and the other with mean age of 6.7 Gyr are the young, low-mass, metal-poor set of blue populations.}
\end{figure*}

\begin{figure*}
\centering
\includegraphics[scale=1.2]{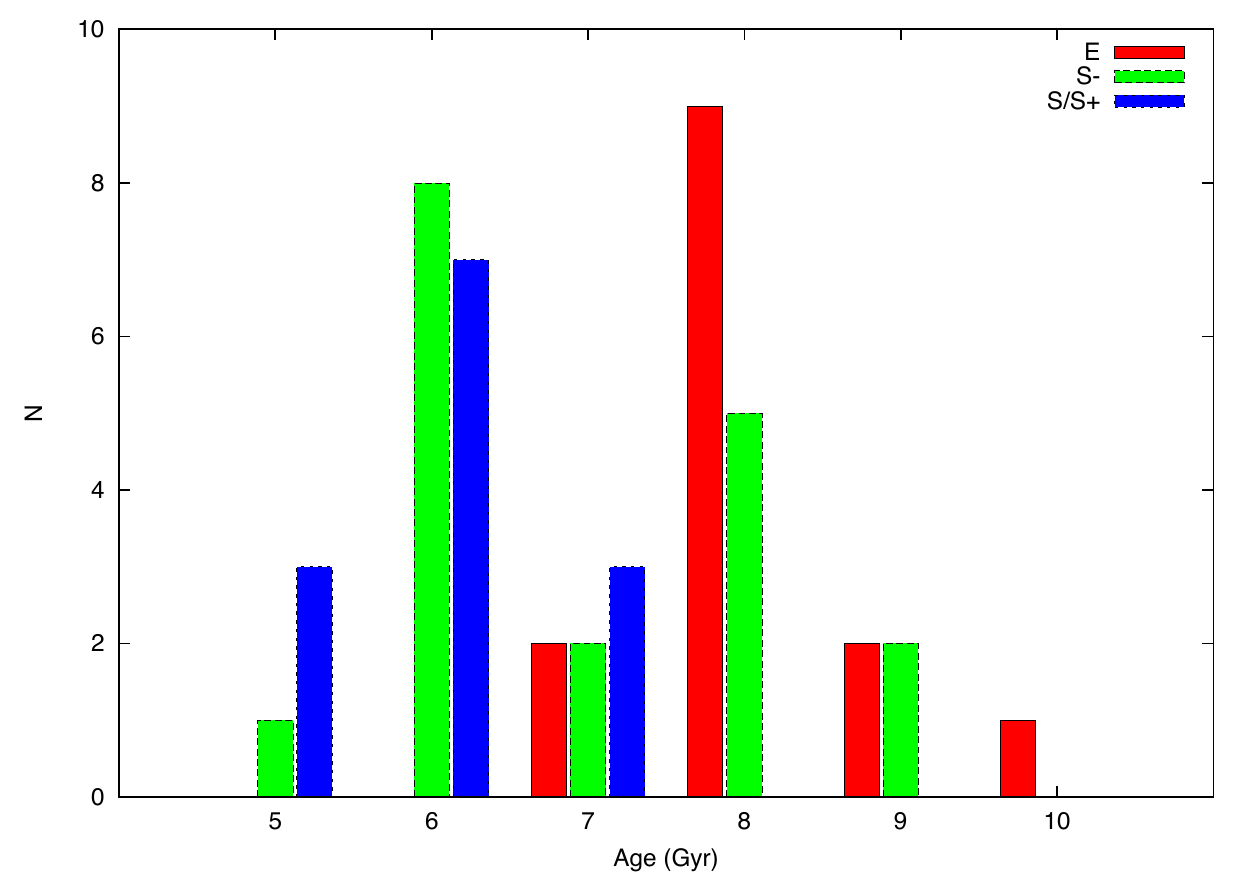}
\caption{Histogram of galaxies of different ages as a function of spectrophotometric galaxy types: E (red), S$-$(green), S (blue). The S and the S+ objects are combined and represented as S type. A clear bimodality of old and young galaxy populations are observed at 6 and 8 Gyr.}
\end{figure*}

\begin{flushleft}
\begin{figure}
\includegraphics[scale=0.7]{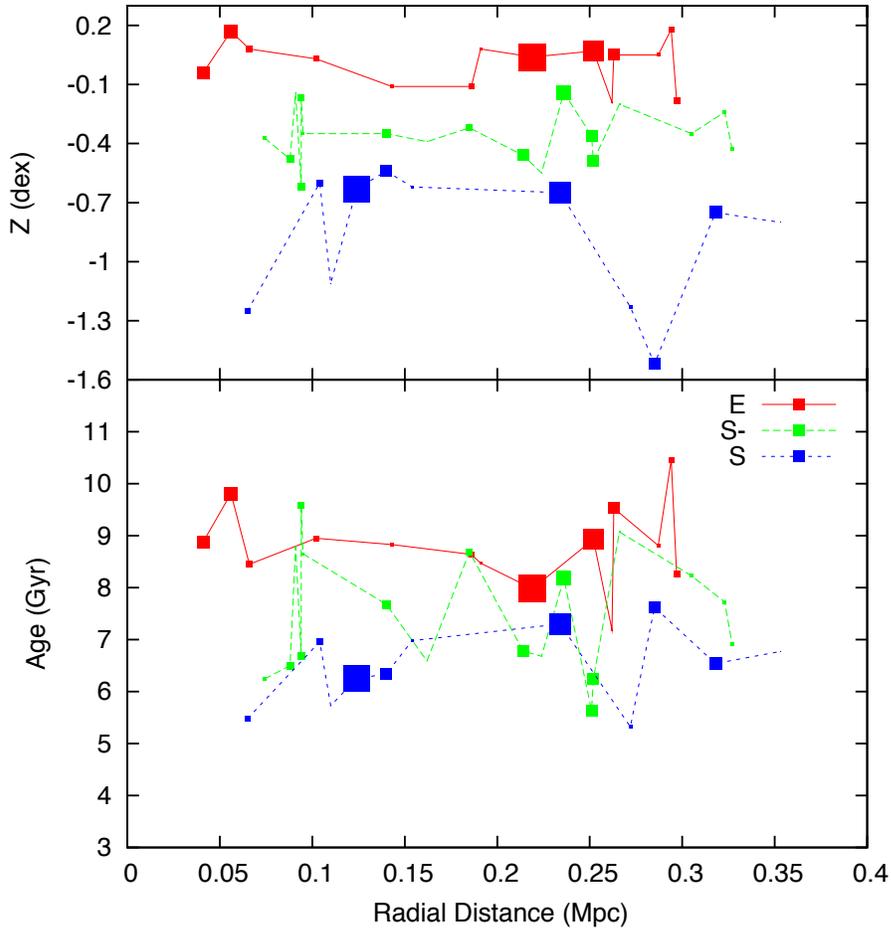}
\caption{Radial change in galaxy age and metallicity as a function of spectrophotometric class and star formation history. The random distribution of galaxies shows no trend in separation by age and metallicity from the cluster center reflecting poor dynamical evolution of the A779 cluster.}
\end{figure}
\end{flushleft}

\begin{deluxetable}{lc}
\tablecolumns{2}
\tablewidth{0pc}
\tablecaption{A779 X-Ray Properties from the Literature}
\tablehead{
\multicolumn{2}{c}{X-Ray Properties} \\
\colhead{} & \colhead{} 
} 
\startdata
$F_x^a$  &  $<$  8.5 $\times$ 10$^{-11}$ erg  s$^{-1}$ cm$^{-2}$  \\ 
$L_x^b$  &  0.09 $\times$ 10$^{44}$ erg  s$^{-1}$  \\ \
$L_x^c$  &  2 $\pm$ 0.19 $\times$ 10$^{43}$ erg  s$^{-1}$  \\ 
$L_x^d$ &  < 0.09 $\times$ 10$^{43}$ erg  s$^{-1}$  \\ 
$L_x^e$ &  < 1.20 $\times$ 10$^{43}$ erg  s$^{-1}$  \\ 
$T_x^f$  & 1.5 keV   \\ 
$\sigma^g$ & 503$^{100}_{-63}$ km  s$^{-1}$ \\ 
$\sigma_p^h$ & 491$^{40}_{-36}$ km  s$^{-1}$ \\ 
\enddata
\tablerefs{Notes. 
a. Kellogg  et al. (1973) at  2$-$6 keV. 
b. Huchra et al. (1992) at 0.1$-$2.4 keV. 
c. White et al. (1997) Bolometric X-rays luminosities. \\
d. Jones \& Forman (1984) using Einstein (Soft, 0.5$-$3 keV). \\
e. Compilation of Einstein, $EXOSAT$ \& $Ginga$ data from David et al. (1993). \\
f. Einstein estimate of x-ray temperature from White et al. (1997).\\
g. Radial velocity dispersion from Zabludoff et al. (1993). \\
h. Radial velocity dispersion from Hwang \& Lee  (2008). \\
}
\end{deluxetable}

\begin{deluxetable}{lllcccrrr}
\tablecolumns{9}
\tablewidth{0pc}
\tablecaption{The Comparison in Table between the Three Sets of Observations Shows the Narrowband Colors, $M_{5500}$ Brightness and the Age, Metallicity Estimations for All the Six Objects using Two Evolutionary Methods, PCA (using Schulz et al. 2002) and GALEV Models (Kotulla et al. 2009)}
\tablehead{
\colhead{}  &  \multicolumn{3}{c}{Narrowband Colors} &   \colhead{}   &
\multicolumn{2}{c}{PCA} & \multicolumn{2}{c}{GALEV} \\
\cline{2-4} \cline{6-7} \cline{7-9}\\
\colhead{Obs.} & \colhead{$(uz$-$vz)$}   & \colhead{$(bz$-$yz)$}   & \colhead{$(vz$-$yz)$}   & \colhead{$M(5500)$}   & \colhead{[Fe/H]}   & \colhead{Age}   & \colhead{[Fe/H]}   & \colhead{Age}\\
\colhead{}   & \colhead{}   & \colhead{}   & \colhead{}   & \colhead{}   & \colhead{(dex)}   & \colhead{(Gyr)}   & \colhead{(dex)}   & \colhead{(Gyr)}} 
\startdata
 & Obj.1. &  &  &  &  &\\
INT & 0.611 & 0.327 & 0.642 & $-$20.164 & $-$0.239 & 8.210 & $-$0.3 & 8.680 \\ 
90 P 2005 & 0.712 & 0.289 & 0.637 & $-$20.266 & $-$0.173 & 7.890 & 0.3 & 8.810 \\ 
90 P 2009 & 0.599 & 0.324 & 0.636 & $-$20.312 & $-$0.316 & 8.690 & 0.3 & 8.400 \\
\hline 
Mean & 0.641 & 0.313 & 0.638 & & $-$0.243 & 8.263 & 0.100 & 8.630 \\ 
 &  &  &  &  &  &  &  \\ 
 & Obj.2.  &  &  &  &  &  \\ 
INT & 0.742 & 0.366 & 0.690 & $-$20.519 & $-$0.017 & 9.350 & 0 & 9.050 \\ 
90 P 2005 & 0.803 & 0.336 & 0.751 & $-$20.613 & 0.028 & 9.63 & 0 & 9.380 \\ 
90 P 2009 & 0.664 & 0.376 & 0.769 & $-$20.687 & 0.048 & 8.810 & 0.3 & 8.570 \\
\hline 
Mean & 0.736 & 0.359 & 0.737 & & 0.020 & 9.263 & 0.100 & 9.000 \\  
 &  &  &   &  &  &  &  &  \\ 
 & Obj.3. &  &  &  &  &  &  \\ 
INT & 0.787 & 0.382 & 0.662 & $-$19.395 & 0.055 & 8.700 & 0 & 9.180 \\ 
90 P 2005 & 0.883 & 0.332 & 0.689 & $-$19.637 & 0.079 & 8.700 & 0.3 & 9.810 \\ 
90 P 2009 & 0.658 & 0.363 & 0.740 & $-$19.680 & 0.041 & 7.980 & 0.3 & 8.550 \\ 
\hline 
Mean & 0.776 & 0.359 & 0.697 & & 0.058 & 8.460 & 0.200 & 9.180 \\ 
 &  &  &  &  &  &  &  &  \\
\cutinhead{More Data} 
 & Obj. 4. &  &  &  &  &  &  \\ 
INT & 0.750 & 0.370 & 0.726 & $-$19.276 & 0.032 & 9.290 & 0 & 9.200 \\ 
90 P 2005 & 0.836 & 0.326 & 0.753 & $-$19.379 & 0.082 & 8.960 & 0.3 & 9.390 \\ 
90 P 2009 & 0.653 & 0.374 & 0.791 & $-$19.449 & 0.076 & 8.470 & 0.3 & 8.560 \\
\hline 
Mean & 0.746 & 0.357 & 0.757 & & 0.063 & 8.907 & 0.200 & 9.050 \\  
 &  &  &  &  &  &  &  &  \\ 
 & Obj. 5. &  &  &  &  &  &  \\ 
INT & 0.742 & 0.379 & 0.735 & $-$20.070 & 0.035 & 9.510 & 0 & 9.180 \\ 
90 P 2005 & 0.901 & 0.348 & 0.724 & $-$20.202 & 0.168 & 9.170 & 0 & 9.510 \\ 
90 P 2009 & 0.679 & 0.385 & 0.767 & $-$20.254 & 0.068 & 8.930 & 0.3 & 8.600 \\ 
\hline 
Mean & 0.774 & 0.371 & 0.742 & & 0.090 & 9.203 & 0.100 & 9.097 \\ 
 &  &  &  &  &  &  &  &  \\ 
 & Obj. 6. &  &  &  &  &  &  \\ 
INT & 0.736 & 0.370 & 0.746 & $-$19.853 & 0.072 & 8.630 & 0 & 9.170 \\ 
90 P 2005 & 0.880 & 0.325 & 0.724 & $-$19.939 & 0.072 & 9.120 & 0.3 & 9.470 \\ 
90 P 2009 & 0.645 & 0.374 & 0.783 & $-$20.007 & 0.030 & 8.950 & 0.3 & 8.550 \\ 
\hline 
Mean & 0.754 & 0.356 & 0.751 & & 0.058 & 8.900 & 0.200 & 9.063 \\ 
\enddata
\end{deluxetable}

\begin{deluxetable}{llllcccc}
\tablecolumns{8}
\tablewidth{0pc}
\tablecaption{The Table Compares the Mean and the Standard Deviation for the Observed Colors from the Three Sets of Observations and for the Measured Ages and Metallicities using the PCA and GALEV Models.}
\tablehead{
\colhead{} & \multicolumn{3}{l}{Narrowband Colors} & \multicolumn{2}{l}{PCA} & \multicolumn{2}{l}{GALEV} \\ 
\cline{2-4} \cline{5-6} \cline{7-8}\\
\colhead{}&\multicolumn{1}{l}{$\sigma(uz-vz)$} & \multicolumn{1}{l}{$\sigma(bz-yz)$} & \multicolumn{1}{l}{$\sigma(vz-yz)$} & \multicolumn{1}{l}{$\sigma$[Fe/H]} & \multicolumn{1}{l}{$\sigma$Age} & \multicolumn{1}{l}{$\sigma$[Fe/H]} & \multicolumn{1}{l}{$\sigma$Age} \\} 
\startdata
Obj. 1 & 0.06 & 0.02 & 0.00 & 0.07 & 0.40 & 0.35 & 0.21 \\ 
Obj. 2 & 0.07 & 0.02 & 0.04 & 0.03 & 0.42 & 0.17 & 0.41 \\ 
Obj. 3 & 0.11 & 0.03 & 0.04 & 0.02 & 0.42 & 0.17 & 0.63 \\ 
Obj. 4 & 0.09 & 0.03 & 0.03 & 0.03 & 0.41 & 0.17 & 0.43 \\ 
Obj. 5 & 0.11 & 0.02 & 0.02 & 0.07 & 0.29 & 0.17 & 0.46 \\ 
Obj. 6 & 0.12 & 0.03 & 0.03 & 0.02 & 0.25 & 0.17 & 0.47 \\ 
\multicolumn{1}{l}{} & \multicolumn{1}{l}{} & \multicolumn{1}{l}{} & \multicolumn{1}{l}{} & \multicolumn{1}{l}{} & \multicolumn{1}{l}{} & \multicolumn{1}{l}{} & \multicolumn{1}{l}{} \\
\hline 
Mean & 0.09 & 0.02 & 0.03 & 0.04 & 0.36 & 0.2 & 0.44 \\ 
\enddata
\tablenotetext{\bf{Notes.}}{The close similarity of the mean and the low standard deviation for colors, ages, and metallicities indicate their fair repeatability and reproducibility by the two photometric systems and the two age$-$metallicity estimation methods.}
\end{deluxetable}
\begin{deluxetable}{lllcrrr}
\tabletypesize{\small}
%\rotate
\tablecolumns{7}
\tablewidth{0pc}
\tablecaption{The Table Lists the A779 Galaxy names, R.A., decl., Photometric Errors for \textit{(uz-vz)}, \textit{(vz-yz)}, and \textit{(bz-yz)}}
\tablehead{
 \multicolumn{1}{c}{No.} & \multicolumn{1}{c}{Galaxy Names} & \multicolumn{1}{c}{RA} & \multicolumn{1}{c}{dDecl.} & \multicolumn{1}{c}{$\pm(uz-vz)$} & \multicolumn{1}{c}{$\pm(vz-yz)$} & \multicolumn{1}{c}{$\pm(bz-yz)$}\\ 
 \colhead{}   & \colhead{}   & \colhead{}   & \colhead{} & \colhead{(mag)}   & \colhead{(mag)}   & \colhead{(mag)}\\}
%\multicolumn{1}{l}{} & \multicolumn{1}{c}{} & \multicolumn{1}{c}{} & \multicolumn{1}{c}{} & \multicolumn{1}{c}{dex} & \multicolumn{1}{c}{Gyr} & \multicolumn{1}{c}{} & \multicolumn{1}{c}{kpc} &  \multicolumn{1}{c}{} & \multicolumn{1}{c}{}}\\ 
\startdata
1 & NGC 2827 & 09:19:19 & 33:52:51 & 0.003 & 0.004 & 0.003 \\ 
2 & SDSS J091920.23+335110.6 & 09:19:20 & 33:51:11 & 0.053 & 0.037 & 0.221 \\ 
3 & NGC 2825 & 09:19:22 & 33:44:34 & 0.001 & 0.002 & 0.002 \\ 
4 & NGC 2826 & 09:19:24 & 33:37:26 & 0.001 & 0.002 & 0.002 \\ 
5 & SDSS J091927.47+334808.4 & 09:19:27 & 33:48:07 & 0.115 & 0.110 & 0.060 \\ 
6 & SDSS J091929.94+335825.9 & 09:19:29 & 33:58:25 & 0.030 & 0.026 & 0.022 \\ 
7 & NGC 2829 & 09:19:30 & 33:38:54 & 0.062 & 0.006 & 0.005 \\ 
8 & SDSS J091930.47+334323 & 09:19:30 & 33:43:23 & 0.152 & 0.141 & 0.079 \\ 
9 & SDSS J091932.98+335553.4 & 09:19:32 & 33:55:52 & 0.122 & 0.109 & 0.075 \\ 
10 & NGC 2828 & 09:19:34 & 33:53:17 & 0.003 & 0.003 & 0.002 \\ 
11 & SDSS J091935.79+334346.6 & 09:19:35 & 33:43:46 & 0.021 & 0.019 & 0.020 \\ 
12 & SDSS J091938.35+334823.8 & 09:19:38 & 33:48:24 & 0.065 & 0.038 & 0.041 \\ 
13 & SDSS J091939.31+334612.1 & 09:19:39 & 33:46:11 & 0.069 & 0.042 & 0.044 \\ 
14 & 2MASX J09193944+3357130 & 09:19:39 & 33:57:12 & 0.008 & 0.006 & 0.006 \\ 
15 & NGC 2830 & 09:19:41 & 33:44:17 & 0.001 & 0.002 & 0.002 \\ 
16 & SDSS J091941.69+334820 & 09:19:41 & 33:48:21 & 0.022 & 0.019 & 0.022 \\ 
17 & SDSS J091942.21+334137.8 & 09:19:42 & 33:41:38 & 0.019 & 0.014 & 0.015 \\ 
18 & SDSS J091942.98 + 334839.5 & 09:19:42 & 33:48:39 & 0.102 & 0.006 & 0.006 \\ 
19 & SDSS J091943.54+333656.5 & 09:19:43 & 33:36:57 & 0.028 & 0.026 & 0.019 \\ 
20 & NGC 2831 & 09:19:45 & 33:44:42 & 0.001 & 0.001 & 0.001 \\ 
21 & SDSS J09194559+3343124 & 09:19:45 & 33:43:12 & 0.006 & 0.006 & 0.005 \\ 
22 & NGC 2832 & 09:19:46 & 33:44:59 & 0.001 & 0.001 & 0.001 \\ 
23 & 2MASX J0919478+334604 & 09:19:47 & 33:46:05 & 0.011 & 0.009 & 0.009 \\ 
24 & 1200-06282293 & 09:19:48 & 33:43:44 & 0.051 & 0.024 & 0.019 \\ 
25 & 2MASX J09195225+3338584 & 09:19:52 & 33:38:59 & 0.003 & 0.003 & 0.002 \\ 
26 & 1200-06282758 & 09:19:55 & 33:40:51 & 0.033 & 0.023 & 0.016 \\ 
27 & SDSS J091957.23+334047.5 & 09:19:57 & 33:40:47 & 0.141 & 0.016 & 0.014 \\ 
28 & NGC 2833 & 09:19:57 & 33:55:39 & 0.001 & 0.001 & 0.001 \\ 
29 & SDSS J092000.91+334225.3 & 09:20:00 & 33:42:25 & 0.024 & 0.021 & 0.021 \\ 
30 & SDSS J092002.6+335804.8  & 09:20:02 & 33:58:05 & 0.014 & 0.015 & 0.014 \\ 
31 & NGC 2834 & 09:20:02 & 33:42:37 & 0.002 & 0.002 & 0.002 \\ 
32 & SDSS J92004.04+334748.3 & 09:20:04 & 33:47:48 & 0.111 & 0.100 & 0.065 \\ 
33 & SDSS J092005.14+334516.4 & 09:20:05 & 33:45:15 & 0.113 & 0.098 & 0.076 \\ 
34 & 2MASX J09200859+3339424 & 09:20:08 & 33:39:42 & 0.003 & 0.003 & 0.002 \\ 
35 & SDSS J92025.35+335151.6 & 09:20:25 & 33:51:52 & 0.101 & 0.077 & 0.076 \\ 
36 & SDSS J92027.84+335035.2 & 09:20:27 & 33:50:35 & 0.020 & 0.020 & 0.013 \\ 
37 & 2MASX J0920286+333902 & 09:20:28 & 33:39:00 & 0.007 & 0.009 & 0.009 \\ 
38 & SDSS J92030.17+333925.4 & 09:20:30 & 33:39:25 & 0.093 & 0.088 & 0.057 \\ 
39 & NGC 2839 & 09:20:36 & 33:39:03 & 0.001 & 0.002 & 0.002 \\ 
40 & SDSS J92037.04+334222.0 & 09:20:37 & 33:42:21 & 0.011 & 0.010 & 0.009 \\ 
41 & SDSS J092041.84+334339.9 & 09:20:41 & 33:43:40 & 0.069 & 0.022 & 0.016 \\ 
42 & 2MASX J09204457+3346220 & 09:20:44 & 33:46:22 & 0.004 & 0.005 & 0.004 \\ 
43 & MCG +06-21-024 & 09:20:45 & 33:42:16 & 0.004 & 0.004 & 0.003 \\ 
44 & SDSS J092047.1+333953.4 & 09:20:47 & 33:39:53 & 0.035 & 0.032 & 0.024 \\ 
45 & SDSS J092048.04+334046.8 & 09:20:48 & 33:40:47 & 0.007 & 0.006 & 0.005 \\ 
\multicolumn{1}{l}{} & \multicolumn{1}{l}{} & \multicolumn{1}{l}{} & \multicolumn{1}{l}{} &\multicolumn{1}{l}{} & \multicolumn{1}{l}{} & \multicolumn{1}{l}{}\\ 
\multicolumn{1}{l}{} & \multicolumn{1}{l}{} & \multicolumn{1}{l}{} & \multicolumn{1}{l}{} & \multicolumn{1}{l}{} & \multicolumn{1}{l}{} & \multicolumn{1}{l}{}\\ 
%\multicolumn{1}{l}{} & \multicolumn{1}{l}{} & \multicolumn{1}{l}{} & \multicolumn{1}{l}{} & \multicolumn{1}{l}{} & \multicolumn{1}{l}{} & \multicolumn{1}{l}{} & \multicolumn{1}{l}{} &  &  \\ 

\enddata
\end{deluxetable}

\end{document}